%% file: Arvix-Main.tex
\documentclass{article}

\usepackage{PRIMEarxiv}

\usepackage[utf8]{inputenc} % allow utf-8 input
\usepackage[T1]{fontenc}    % use 8-bit T1 fonts
\usepackage{hyperref}       % hyperlinks
\usepackage{url}            % simple URL typesetting
\usepackage{booktabs}       % professional-quality tables
\usepackage{amsfonts}       % blackboard math symbols
\usepackage{nicefrac}       % compact symbols for 1/2, etc.
\usepackage{microtype}      % microtypography
\usepackage{lipsum}
\usepackage{fancyhdr}       % header
\usepackage{graphicx}       % graphics
\graphicspath{{media/}}     % organize your images and other figures under media/ folder
\input{commands}
\gpcefalse

\title{IsiSPL: Toward an Automated Reactive Approach \\to build Software Product Lines
\thanks{\textit{\underline{Citation}}: 
\textbf{Authors. Title. Pages.... DOI:000000/11111.}} 
}

\author{
  Hlad Nicolas \\
  LIRMM \\
  University of Montpellier \\
  CNRS\\
  Montpellier, France\\
  \texttt{hlad@lirmm.fr} \\
  \and
  Seriai Abdelhak-Djamel \\
  LIRMM \\
  University of Montpellier \\
  CNRS\\
  Montpellier, France\\
  \texttt{seriai@lirmm.fr} \\
     \and
  Dony Christophe \\
  LIRMM \\
  University of Montpellier \\
  CNRS\\
  Montpellier, France\\
  \texttt{dony@lirmm.fr} \\  
}

\begin{document}

\maketitle

\begin{abstract}
\input{Abstract/abstract-v2}
\end{abstract}

% keywords can be removed

\input{Introduction/introduction-v6}

\input{Overview/overview-v9}

\input{Integration/integration-v2}

\input{Generation/generation-v3}

\input{Validation/validation-v2}

\input{RelatedWork/relatedWork}

\input{Conclusion/conclusion-en}

\section*{Acknowledgments}
This work was supported by the \textit{ISIA Group}. \url{www.groupe-isia.com}

%Bibliography
%\bibliographystyle{unsrt}  
%\bibliography{reference}

\input{./biblio}
\end{document}

%% file: commands.tex
\newif\ifgpce

\usepackage[ruled,vlined, linesnumbered]{algorithm2e}

\makeatletter
\newcommand{\thickhline}{%
    \noalign {\ifnum 0=`}\fi \hrule height 2pt
    \futurelet \reserved@a \@xhline
}

%% file: Abstract/abstract-v2.tex
Over the past decades, Software Product Lines (SPLs) have demonstrated the benefits of systematic reuse to increase software development productivity and software product quality. 
Of the three adoption approaches, i.e. extractive, proactive and reactive, the reactive approach seems the most suitable for software development in practice.
The strength of this approach is that it remains close to classical software development practices. 
In fact, it avoids a complete analysis of the business domain and its variability (i.e. proactive approach),  
and avoids requiring a set of product variants that covers this domain (i.e. extractive approach).
Nevertheless, despite these advantages, we believe that the main obstacle of the reactive approach adoption is the lack of automation of its re-engineering process.

This paper proposes isiSPL: a reactive-based approach that facilitates both construction and maintenance of an SPL.
The construction of the SPL is based on a quasi-automatic process. 
The maintenance of the SPL can be made on a white-box SPL implementation, generated by isiSPL. 
isiSPL is based on two steps: first, the identification and integration of the artefacts of a newly created product into the structure of the SPL;
second, the selection of a set of artefacts and their composition to generate a new product that can either partially or completely meet the requirements provided for a product intended by a developer. 
We have implemented isiSPL and validated its integration and generation using the two different sets of products from ArgoUML-SPL and Soduko-SPL. 

%% file: Introduction/introduction-v6.tex
\section{Introduction}

Software Product Line Engineering (SPLE) has shown its advantages in terms of increased productivity and improved product quality based on systematic reuse.
These advantages have been beneficial to many companies as evidenced by the SPLC's hall of fame\footnote{\href{https://web.archive.org/web/20210713201727/https://splc.net/fame.html}{https://splc.net/fame.html}}.
Nevertheless, it is clear that despite these many advantages, SPLE is far from being widely adopted by the software development community.
We believe that the reasons behind this low adoption of SPL are mainly related to the difficulties of developing and maintaining an SPL.
In the literature, three approaches to constructing product lines are proposed: proactive, extractive and reactive \cite{pohl_software_2005}.
Each of these three approaches has its own weaknesses which lead to one or more obstacles to SPL adoption.

The proactive approach \cite{eriksson_introduction_nodate,pohl_software_2005} is based on a thorough analysis of the business domain of the product line to identify and build all of its constituents. 
It is often referred to as a "from scratch" process.
First, the developers of the SPL identify the \emph{features}, i.e. a set of distinguishable and reusable product functionalities.
Then, they build an SPL implementation that gathers all the specific and shared features, i.e. implemented by a set of artefacts (e.g., source code, resources files, tests...) and which must take into account the variability of the related business domain \cite{gacek_implementing_2001}.
This approach is adopted mainly by large software editors since they have a clear vision of their business field outline, and thus of the products they want to develop.

The extractive approach is based on a \emph{Bottom-Up} process to extract the constituents of an SPL from an already existing set of ad-hoc product variants \cite{krueger_easing_2002}. The advantages are that the SPL's constituents can be recovered (e.g. artefacts, features, Feature-Model\cite{batory_feature_2005}) by analyzing source codes and other artefacts associated with these products (e.g. software requirement, user manual, design documents...).
Therefore, as this process is usually fully automatic, the cost of building an SPL based on this process is usually low.
However, the obstacle to the massive adoption of this approach lies in its prerequisite: the availability of a large set of product variants already developed in an ad-hoc manner that covers all the SPL business-domain features in question.
Many companies cannot reach this prerequisite.
In addition to this limitation, the existing works that implement an extractive approach fail to propose a white-box SPL implementation \cite{linsbauer_recovering_2013, ziadi_towards_2014, al-msiedeen_reverse_2014}, since the SPL is often intended to be used by product-managers to configure its products, and not by developers to maintain and evolve the SPL if necessary.
According to Lehmann's first law of software evolution, any software that does not evolve will die \cite{DBLP:conf/ewspt/Lehman96a}.

\iffalse
The reactive approach is based on an iterative process that gradually builds an SPL \cite{buhrdorf_salions_2003}.
Whenever a new product is to be created, the SPL is re-engineered to take into account that new product.
In some ways, the reactive approach is similar to a \textit{classic} software development process (see $(a)$ and $(b)$ in Figure \ref{fig:reactive}).
The core of this approach is about integrating missing new features and their artefacts into an SPL.
Some features and artefacts of this new product may already exist in the current version of the SPL.
Once this integration has been carried out, the new product can thus be generated.
It is therefore the same domain analysis as carried out for the proactive approach but here reduced to one product at a time.
The construction of the line is considered finished if all the products created and integrated in the SPL cover all the features of the business domain and their composition constraints.
\fi

\begin{figure}
    \centering
    \ifgpce
    \includegraphics[width=0.43\textwidth]{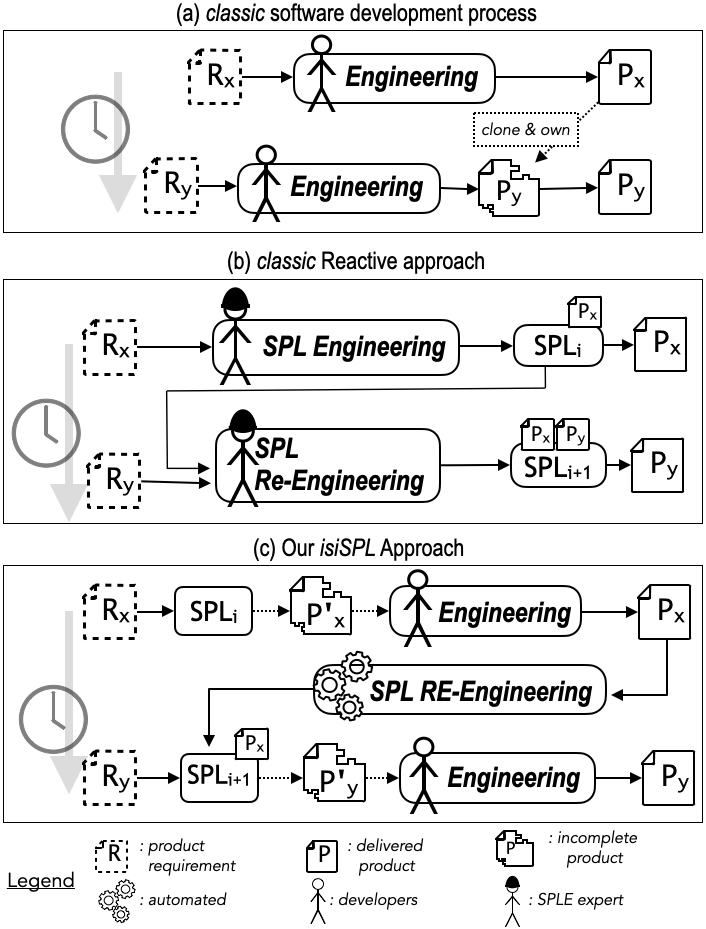}
    \else
    \fbox{\includegraphics[width=0.5\textwidth]{Introduction/approaches.png}}
    \fi

    \caption{A side by side comparison of $(a)$ the classic software development, $(b)$ the reactive approach, and $(c)$ our proposed \textit{isiSPL} approach.}
    \label{fig:reactive}
\end{figure}

Finally, the reactive approach can be described as an iterative process that gradually builds an SPL \cite{buhrdorf_salions_2003}. 
Rather than engineering an entire SPL in one iteration, the reactive approach spreads the SPL investment over time, since it is iteratively built product after product. 
In some ways, this approach is comparable to a \textit{classic} software development process (see $(a)$ and $(b)$ in Figure \ref{fig:reactive}):
both require developers to engineer their new products, starting from requirements $R_x$ (i.e. a description of a new product formulated by a customer) that they receive over time.
However, the reactive approach replaces the tedious \textit{clone\&own} \cite{Zhang_Peng_Xing_Zhao_2012} often found in classical software development (see $(a)$), by a systematic reuse through the SPL. 
Since this approach focuses primarily on the delivery of new products, it benefits companies working on \emph{custom software}, who look for a rapid return of investment from their SPL, or companies whose business outline is still uncertain.

Some studies have reported the industrial potential of the reactive approach \cite{buhrdorf_salions_2003, ghanam_reactive_2010}.
However, others report that despite its advantages over the proactive and the extractive, the reactive approach still suffers major difficulties related to the iterative re-engineering of an SPL \cite{Krueger_Berger_2020, Akesson_Nilsson_Krueger_Berger_2019, hlad_facilitating_2019}.

Indeed, the problem is that the integration of new requirements is difficult and requires expertise related to the business domain and the technical domain.
For the business expertise, which is the role of product-managers, it is a question of identifying all features already implemented in the SPL, the new features destined for a new product, and the variability relationship between them.
For the technical expertise, which is the developers' role, it is about modifying the existing SPL implementation and integrating the artefacts of the new required features without jeopardizing the SPL ability to re-generate its older products.
Since its set of products increases over time, an SPL becomes more complex to re-engineer with more features, more features interactions, and richer variability.
Furthermore, developers have to carefully measure the impacts of their implementation choices, since they may have an unpredictable effect on the next iterations.
In some cases, the re-engineering effort, and thus its cost, can be so high that it becomes a risk for the company \cite{schmid_economic_2002}.
Therefore, this difficulty of manually building and evolving an SPL is seen as the main obstacle towards widespread adoption of the reactive approach.

In this paper, we present our contribution to addressing the SPL re-engineering difficulties in the reactive approach.
We propose a reactive-based approach called \textit{isiSPL}, for automating most of the SPL engineering. 
In our approach, depicted in Figure \ref{fig:reactive} $(c)$, we retain the same principle of product development as the reactive approach $(b)$, where new product requirements $R_x$ drive the incremental construction of an SPL. 

We propose to reduce the difficulties of re-engineering by transforming it into a classical development step for developers. In order to speed up development, developers generate a partial product $P'_x$ from $SPL_i$, and complete it (if needed) \textit{outside} of an SPLE environment by designing the features required by $R_x$.
This gives the greatest flexibility to build  SPL products by experts and non-experts alike.
Once $P_x$ is obtained, isiSPL automatically integrates it into $SPL_i$ to obtain $SPL_{i+1}$.   
Thus, the features introduced by $P_x$ will be reusable in future iterations, saving time during the development of new products $P_y$. 

Finally, unlike other existing automated approaches \cite{linsbauer_recovering_2013, ziadi_towards_2014, al-msiedeen_reverse_2014}, isiSPL gives developers the possibility to directly manipulate a \emph{white-box} SPL implementation. 
Thus it ensures full adherence to Lehmann's first law \cite{DBLP:conf/ewspt/Lehman96a}.
We propose this SPL implementation to be an annotated source code, since it is known to be intuitive for most developers \cite{kastner_integrating_nodate}.

%%%%%%
\iffalse
In this paper, we present an approach to automate the reactive construction of an SPL.
We propose an adaptation and a materialization of the generic process so that this automation is possible.
In our approach called IsiSPL, depicted in $(c)$ of figure \ref{fig:reactive}, the creation of a new product can be achieved partially or completely by its generation from the existing version of the SPL.
The generated product is complete if all its required features are already available and if only their implementations (and no other features) are included in this generated product.
Otherwise, it is partial.
If the product is partially generated, the developer supplements it by adding, modifying or removing artefacts.
The SPL implementation takes the form of annotated code, which is known to be intuitive for most developers.
This gives the greatest flexibility to build the SPL by experts or non-experts.
All products that are really new, i.e. that could not be completely generated from the current version of the line, are analyzed to identify their artefacts and locate their features.
Then, these artefacts and features are automatically integrated into the current version of the SPL, to make it evolve and obtain a new version where the generation of this latest product will be possible.
\fi

Our paper is organized as follows:
Section \ref{sec:overview} presents in detail isiSPL and the problem linked to the automation of the SPL engineering.
Section \ref{sec:integration} presents the integration of new products and an the SPL evolution process.
Section \ref{sec:generation} presents the isiSPL Generation phase, which creates an annotated code of the SPL implementation.
Section \ref{sec:validation} presents the validation of an isiSPL prototype, its results and threats of validity.
Section \ref{sec:relatedwork} refers to related works and the final section concludes the paper.

%% file: Overview/overview-v9.tex
\section{Goal And Problem Statement}
\label{sec:overview}

\subsection{IsISPL, towards an automated reactive approach}
\label{sec:isiSPL}

\begin{figure}
    \centering
    \ifgpce
    \includegraphics[width=0.463\textwidth]{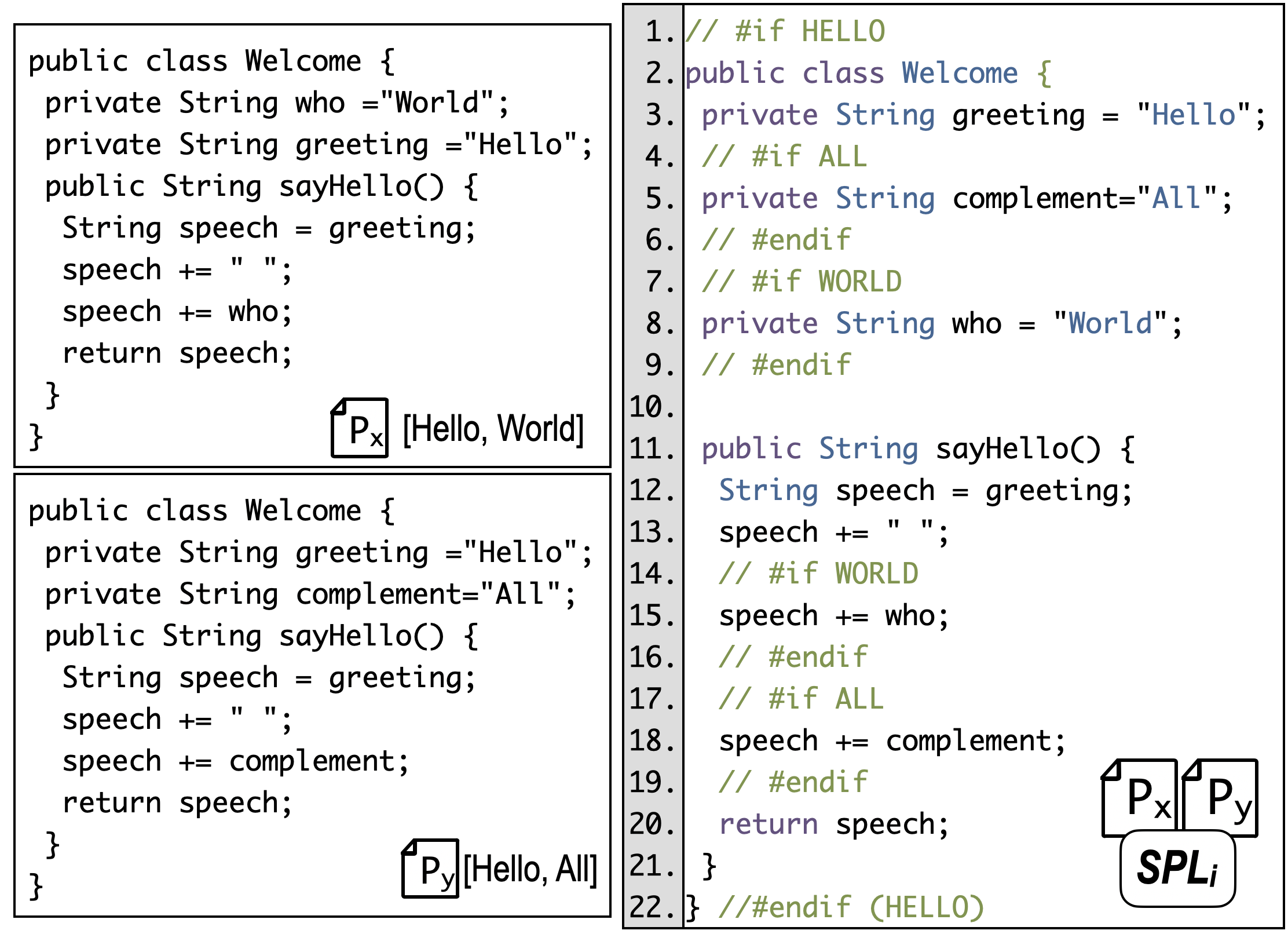}
    \else
    \fbox{\includegraphics[width=0.7\textwidth]{Overview/integration-example.png}}
    \fi
    \caption{An example of two products being integrated by IsiSPL.}
    \label{fig:integration-example}
\end{figure}

\begin{figure}
    \centering
    \ifgpce
    \includegraphics[width=0.415\textwidth]{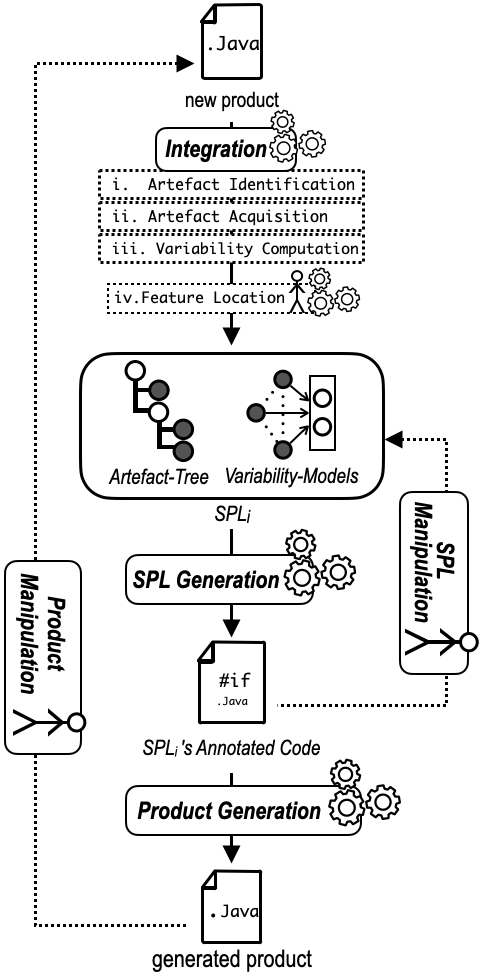}
    \else
    \fbox{\includegraphics[width=0.7\textwidth]{Overview/semi-auto-overview-v2.png}}
    \fi
    \caption{Overview of isiSPL}
    \label{fig:overview}
\end{figure}

Our goal is to help automate the various stages of the reactive process of building a software product line. 
We have defined and configured IsISPL, an automatic reactive process for the creation of SPL and the steps of which are shown in Figure \ref{fig:overview}.
Considering the example of the two products shown in Figure 2. $P_x$ is a simple Hello World printer, having two features: Hello (prints Hello) and World (prints World). 
IsiSPL starts when a request is provided to create a new product based on its requirements. 
In our example, the requirements describe a new product $P_y$, which prints "Hello All".
 The first step for the developers is to comprehend the requirements as a set of features that they need to implement or reuse based on their current SPL.
 The \emph{Product Generation} step of the process allows you to create an incomplete product from the selection of a set of features.
 Thus, from the given requirements, developers estimate that $P_y$ introduces the new feature "All" and can reuse the feature "Hello".
 However, since the feature "World" is close to the expected feature "All", they select the two features "World" and "Hello" to generate an incomplete product of $P_y$.
 Within this phase, developers can modify the generated code until it corresponds to the implementation of the requirements (i.e. list of features). 
The obtained product $P_y$ is shown in the left bottom of Figure \ref{fig:integration-example}.
The completed product $P_y$ is used to re-engineer and thus evolve the previous SPL version.

This transition from version $SPL_i$ to $SPL_{i+1}$ is based on two phases: the \emph{Integration} and the \emph{SPL Generation}. 
The Integration phase takes as input the source code of the and a list of features, i.e. their name and their description, that the developer implements. 
In our example, this corresponds to the code of $P_y$ and the two feature names Hello and All. 
The goal of the Integration phase is to identify the specific new features introduced by a new product and add them inside the SPL correctly. 
This integration is conducted so that adding a new feature will not affect the behavior of previously integrated ones. 
For instance, after integrating the feature "All", it is necessary that the developers are still able to generate the feature "World" inside a product (see the code of $P_x$ and $P_y$ Figure \ref{fig:integration-example}). 

Furthermore, the feature behavior must remain unchanged, e.g., when they reuse feature "All", the product does print "All".
A feature behavior depends on its actual implementation inside the product.
A feature is implemented by various artefacts, such as source code, images, script, data, etc.
We consider here artefacts that can be found in the source code of a product.
They can be class declarations, fields, methods, statements, import declarations, etc.
In order to correctly acquire the feature "All", the Integration phase identifies all artefacts inside the code of $P_y$.
Then, it acquires only the new artefacts into a structure called the \emph{Artefact-Tree} which takes the form of a common tree structure for all the source code artefacts ever integrated in the SPL.

Next, the Integration phase computes the variability of the SPL based on the artefact variability it finds inside the products it has integrated.
From our previous example, the class $Welcome$ appears in both product $P_x$ and $P_y$.
However, these classes have different implementation in $P_x$ and $P_y$.
These two versions of the class $Welcome$ share some fields and the same method but have two different implementations.
From this analysis, the Integration phase generates the variability models and traces the artefacts to their corresponding feature (i.e. Feature Location).
Having done evolve the SPL based on the integration of $P_y$ artefacts, the Integration phase ends, and the \emph{SPL Generation} phase starts.
The SPL Generation aims to create the annotative representation of the SPL implementation.
We obtain the annotated code of $SPL_i$, on the right side of Figure 2.
The developers can directly modify this representation inside an \emph{SPL Manipulation} phase. 
Thus it permits the developers to operate changes on the SPL itself if necessary (evolving the SPL).

\subsection{The Artefact Granularity}
\label{sec:granularity}

\begin{figure}
    \centering
    \ifgpce
    \includegraphics[width=0.463\textwidth]{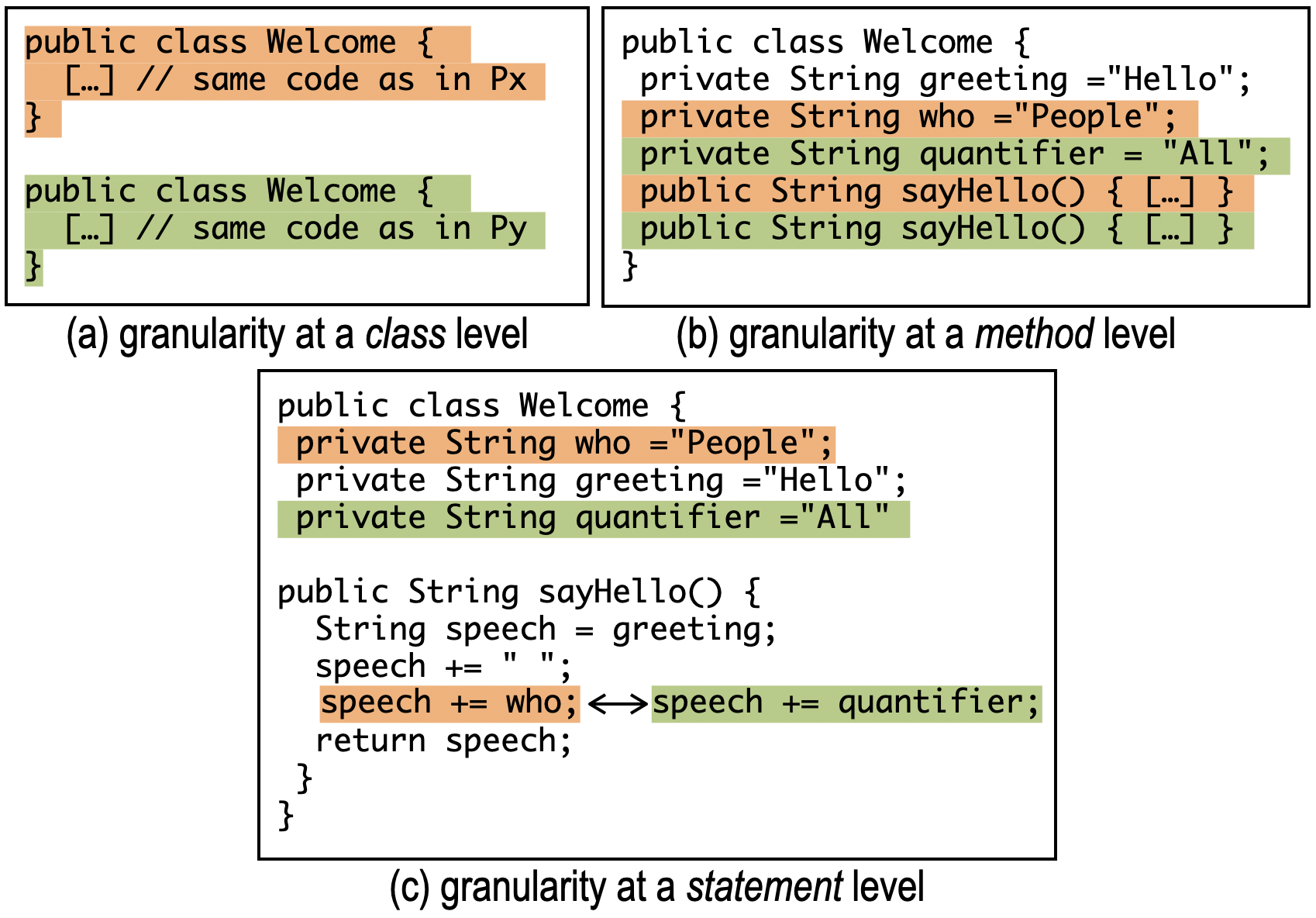}
    \else
    \fbox{\includegraphics[width=0.6\textwidth]{Overview/example-granularity.png}}

    \fi
    \caption{An example of how the artefact granularity can affect the integration of our example (Fig. \ref{fig:integration-example}). In orange are the artefacts from $P_x$ and in green the ones from $P_y$.}
    \label{fig:example-granularity}
\end{figure}

K\"aster et al. \cite{kastner_granularity_2008} indicated that the granularity of the artefact has a determining role over the construction of an SPL, its comprehension by the developers, and up to the product generation.
We completely agree with this assertion.
For this reason, in IsiSPL, we have chosen to consider artefacts at a fine level, that of source code statements and not at a higher level like that of methods or classes (see figure \ref{fig:example-granularity}). 
%The consequence of defining the artefacts at the code statement level:
This has the advantages of enabling us to :
i) to capture the commonality and variability of products at a very fine level, that of statements.
This makes it possible to generate products that are closest to the requirements provided by composing existing artefacts.
Consequently, this makes it possible to reduce the \emph{distance} between the products expected by the developers and the ones obtain from the Product Generation phase, and thus to reduce the effort of developers during the Product Manipulation phase;
ii) to accurately distinguish the implementation of each feature at the finest level possible. This has the consequence of enhancing the overall SPL comprehension;
and iii) to factorize more of the code, i.e. to improve the maintainability of the SPL. 

However, as we see in $(c)$ of Figure \ref{fig:example-granularity}, studying statements as individual artefacts makes us, for instance, consider their position inside the methods, which is essential to preserve the feature behaviors. 
Thus, in contrast with the previously mentioned advantages, this granularity level raised problems that we expose in the rest of this section.

\subsection{Problems related to the Automation}

% explication possiblement à revoir
\subsubsection{Problem 1: Avoiding the duplication of feature implementations}
When we integrate artefacts to an SPL implementation, we must ensure that we do not add artefacts that are already integrated and thus avoid adding the same feature implementation. 
This is to avoid the undesirable effect of maintaining two versions of a same feature inside an SPL implementation.
Therefore, it requires each artefact to be unique within an SPL implementation.
This unicity has to be guaranteed during the acquisition of new artefacts, by comparing the artefacts from a new product with those already inside the SPL. 
For instance, classes, methods and field artefacts are easily distinguishable because they are only declared once per file.
However, with our fine granularity goal, the artefact unicity has to consider the particular case of statements that can appear multiple times inside a product.
Thus, the first problem is to retain the unicity of all the artefacts during the SPL re-engineering, i.e. the Integration phase.
Thus the first question is \textbf{$Q_1$: How do we ensure the unicity of the artefacts within an SPL implementation?} 

\subsubsection{Problem 2: Preserving features behavior}
The Integration has to preserve the behavior of each feature, newly or previously integrated into the SPL. 
These behaviors are implemented by a set of artefacts, which can be statements. 
Furthermore, the order of the statements impacts the feature behavior.
Therefore to preserve it, the SPL implementation must preserve the sequence of statements, while preserving the unicity of each artefact (problem 1).
Thus our second question is \textbf{$Q_2$: How do we preserve the sequences of statements, and thus the behavior of an integrated feature?}

\subsubsection{Problem 3: Managing variability}
Every SPL is characterized by its ability to vary in order to generate different products. 
Thus, an SPL must distinguish the common parts from the variable parts in its implementation. 
We refer to those variable parts as \textit{variation points}. 
Knowing the variation points allows the developers to understand the variability in the implementation of their SPL. 
Moreover, it allows them to design their products by selecting the convenient artefacts of a variation point to be included inside a product. 
Thus for a proper SPL maintainability, the developers need to know where those variation points are and which variable artefacts can be selected. 

However, a binary vision of variability, i.e. \textit{common} and \textit{variable}, is insufficient.
Developers also need to know the \textit{constraints} among the feature (and thus among artefacts). 
From our example, they need to know that the artefacts of the feature \textit{All} cannot be selected without the feature \textit{Hello}.
Otherwise, their products will be badly generated, as the artefacts of \textit{All} are declared inside the artefacts of \textit{Hello} (see annotated code of Figure \ref{fig:integration-example}).
This requires extracting the necessary variability constraints between the features and between the artefacts. 
Thus the third question is \textbf{$Q_3$: How do we automatically extract the SPL variability and its constraints?}

\subsubsection{Problem 4: Feature Location}
The variability goes hand in hand with the extraction of traces between features and their implementation, as is a set of artefacts.
It is essential that each artefact can be included in a new product in order to be reused.
If product generation is done using a feature selection, each artefact must be linked to at least one feature or feature interaction.
For example in the SPL implementation of Fig. \ref{fig:integration-example}, the feature \textit{World} is implemented by the artefacts line 8 and 15, and \textit{All} by the ones line 5 and 18.
This sets the fourth problem of the Feature Location (\cite{dit_feature_2013}) during the product integration.
\textbf{$Q_4$: How to extract traces between features and artefacts during product integration?} 

\subsubsection{Problem 5: Representing an SPL implementation}
As we explain previously, developers need to be able to manipulate their SPL implementation.
For this purpose, isiSPL generates a representation of an SPL implementation as an annotated code.
However, the annotation-based representation suffers from a readability problem when many annotations appear in the code.
Annotated code can become unreadable and thus frustrating to handle for developers.
This is often referred to as the \textit{if/else hell} problem \cite{aleixo_comparative_2012}. 
\textbf{$Q_5$: How to minimize or avoid the \textit{if/else hell} problem when generating the annotated-code of an SPL?}  

\smallbreak
In the rest of our paper, we present our solutions for the identified problems, except \textit{Problem 4}.
We will not present a specific automated Feature Location technique, leaving it as future works. 
However, we propose some ideas to be explored to automate this particular step.

%% file: Integration/integration-v2.tex
\section{The Integration Phase}
\label{sec:integration}

This section presents the Integration phase and our solutions for its related problems described in section \ref{sec:overview}. 
We present this phase over the integration of a product $P_z$, which implements the features \textit{Hello}, \textit{All} and \textit{People} (see $P_z$ code in the left part of Figure \ref{fig:identification}) 

\subsection{Artefact Identification}
\label{sec:ident}

The artefact identification aims at finding all the artefacts from the source code from a software product. 
Here we mention only the case of Java artefact, but this technique can be applied to any software language with an Abstract Syntax Tree (AST).
For Java, we consider over 65 types of artefacts, such as (e.g.) class declaration, import declaration, field, methods, statements (for, while, if, switch case...), initializer, Java annotation, etc. 

\begin{figure*}
    \centering
    \ifgpce
    \includegraphics[width=.9\textwidth]{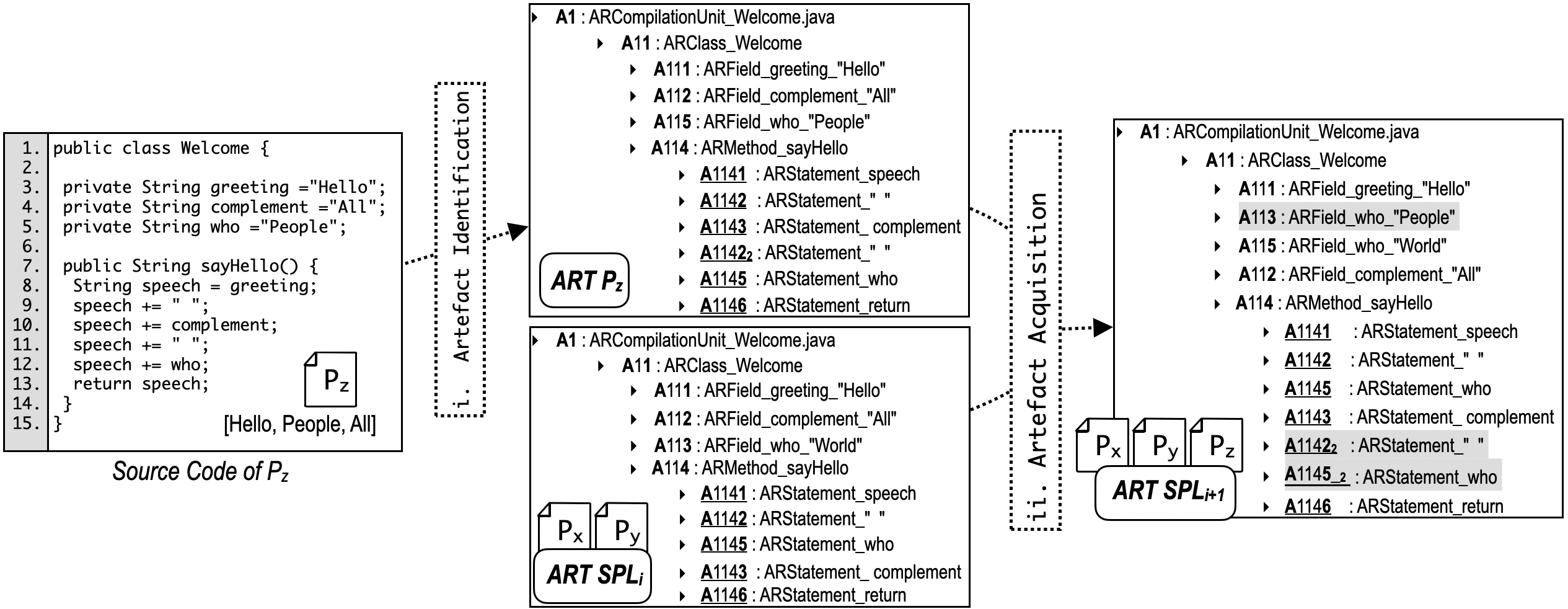}
    \else
    \fbox{\includegraphics[width=1\textwidth]{Integration/identification.png}}
    \fi
    \caption{Identification and acquisition of $P_z$ artefacts inside $SPL_{i}$, to create $SPL_{i+1}$. Artefacts in sequence in both $P_z$ and $SPL_{i}$ ARTs are underlined. All newly added artefacts inside $SPL_{i+1}$ are highlighted in grey.}
    \label{fig:identification}
\end{figure*}

An artefact has a \textit{value}, which represents its associated code.
In Figure \ref{fig:identification}, we run the identification over the code of $P_z$. The identification creates an artefact for each of the main nodes inside the AST of $P_z$. In our example : there is an artefact for compilation unit (\textit{AR}CompilationUnit\_Welcome.java), for the class $Welcome$ (\textit{AR}Class\_Welcome), and so on. 
The identified artefacts are stored inside an Artefact-Tree (ART).
Similarly to an AST, An ART stores the artefacts by following a hierarchy of \textit{parent-children} (see ART of $P_z$ in  Fig. \ref{fig:identification}). 
Thus, an ART looks like the AST of a targeted language, and as such we propose to consider one ART for each file in a product.  

The artefacts all have a different identifier (ID) inside an ART (e.g. $A1$, $A11$, etc., to $A1146$ in the ART of $P_z$).
These identifiers (IDs) are essential to avoid the acquisition of artefacts already integrated into the SPL, and thus avoid problem 1, related to the duplication of feature implementation.
These IDs will allow us to compare the artefacts present in two ARTs.
From this comparison, new artefacts will be acquired by comparing the ARTs inside the SPL with the ones of a product.
To ensure that there is only one instance of each artefact in the SPL, we use IDs to add artefacts of $P_z$ to the SPL which are not already in $SPL_i$. 

An artefact's ID is computed as a hash of the artefact's value, added to the ID of its parent.
This recursive construction of IDs, which starts from the ART root, guarantees that each artefact has a unique ID inside one ART. 
Moreover, we can extend the unicity property of the IDs to the entire product. 
The compilation unit artefact (the root an ART) is given a value that is the relative path of the actual file, inside the product.
Since two different files will necessarily have a different relative path, two ART roots cannot share the same ID.  
With this construction of IDs, only artefacts that share the same parental hierarchy (from the same ART root) and that have the same value, will share the same ID.

For instance, this can happen if two statements of the same method share the same code, like in $P_z$ code at lines 9 and 11. Consequently, the ART of $P_z$ has two artefacts \textit{ARStatement\_"\textvisiblespace"}, with the same ID (\(A1142\)). 
We call these artefacts \textit{twins}.
We introduce the notion of twin artefacts to differentiate tow equals statements (i.e. same code), but at a different place in a same method. 
In our example, the first artefact \textit{ARStatement\_"\textvisiblespace"} adds a \textit{blank} to separate \textit{Hello} and \textit{All}, while the second separates \textit{All} from \textit{People} in the final string. 
To differentiate them, we specifically add another parameter called \textit{twin-id} to construct the ID of each statement artefact.
%This parameter is named "\textit{twin-id}", and is set to the numbers of twins that precede the artefact, plus one. 
This parameter represents the number of twins that precede it plus one, if any.
In our example, this is symbolized by the index given to the artefact \(A1142_2\). 

Finally, we note that the position of an AR-Statement does not intervene during the construction of its ID. 
However, the sequence has to be saved somehow, to maintain the behavior implemented in the methods. 
The sequences are stored in the ART in the same way as they are in an AST. 
We see in the ART of $P_z$ that we recover the order of initial statements for the sequence of $sayHello()$ by visiting the ART in-depth.

\subsection {Artefact Acquisition}
\label{sec:acquisition}

The artefact acquisition aims at adding the identified artifacts introduced by a new product to the SPL.
To do so we merge the $P_z$ ARTs with the $SPL_i$ ARTs to produce a \emph{super-ART} for $SPL_{i+1}$ (see Figure \ref{fig:identification}).
This super-ART contains all the artefacts of both ARTs, without duplicates, as we merge artefacts with the same ID into a single artefact. 
Note that for any new files introduced by the product, their corresponding ARTs are absent from the SPL, thus there can be added to $SPL_{i+1}$ without merging.

With problem 2, we emphasized the necessity to preserve the sequence of artefacts inside the ART. 
Since two ARTs may contain two different statement sequences, their merge must ensure that the output ART preserves them without creating duplicates. 
We propose building a super-sequence covering both sequences from the product ART and the SPL ART, inside a merged ART.
Our solution relies on the \textit{Longest Common Subsequence (LCS)} algorithm \cite{878178}, but a full presentation of LCS is out of the scope of this paper.

For two sequences as input, the LCS algorithm computes the longest sequence of elements that are common to both sequences. 
We present our super-sequence algorithm based on LCS in Algorithm \ref{alg:LCS}.
In our example, the LCS function outputs the $lcs$ sequence $[A1141, A1142, A1143, A1146]$ (line 1).
We initialize the super sequence $supSeq$ from the first element of $S1$, and $S2$, to the first element of $lcs$ (line 2).
For every two consecutive IDs $i1 and i2$ in $lcs$ (line 3), we get the sub-sequence ($subSeq()$) that is between the two IDs, in both ARTs (line 6 and 7); 
e.g. in $SPL_i$ ART, the sub-sequence in-between $A1142$ and $A1143$ is $[A1145]$; while the same interval in $P_z$ ART is empty.
Finally, we add the rest of the artefacts  ($restOf()$) of both ARTs at the end of the $SupSeq$ (line 10).

\begin{algorithm}
\SetAlgoLined
\KwIn{Two sequences  S1, S2}
\KwOut{A super sequence} 
 Sequence lcs $\gets$ LCS(S1, S2)\; 
 Sequence SupSeq $\gets$ S1.subSeq(S1[0], lcs[0]) + S2.subSeq(S2[0], lcs[0])\;
 \ForEach(){consecutive-IDs i1, i2 $\in$ lcs}{
     Sequence seq1, seq2 = null\;
     seq1 $\gets$ S1.subSeq(lcs[i1], lcs[i2])\;
     seq2 $\gets$ S2.subSeq(lcs[i1], lcs[i2])\;
     SupSeq $\gets$ SupSeq + lcs[i1] + s1 + s2\;
 }
\Return{SupSeq + restOf(S1) + restOf(S2)}
 \caption{Our super sequence algorithm based on the \textit{Longest Common Subsequence} (LCS)\label{alg:LCS}}
\end{algorithm}

The result is the super-sequence in the ART $SPL_{i+1}$ on the right side of Fig. \ref{fig:identification}.

However, in some cases, it is mandatory to duplicate artefacts to build a correct super-sequence.
In the ART of $P_x$ and $SPL_i$, the artefacts $A1145$ of the statement $who$ appears in two different positions in the sequence. 
To align both sequences, we had to introduce a duplication of $A1145$ to create the super-sequence in the ART of $SPL_{i+1}$.
These two artefacts correspond to the implementations of two different features. We add as a visual aid the corresponding artefact ID from the ART next to the annotation. 
In order of appearance: the first one is from $P_x$ and is part of the implementation of the feature \textit{World}; the second is from $P_z$ and part of the feature \textit{People}
However, to keep the unicity of our artefacts in $SPL_{i+1}$, i.e. to distinguish both artefacts $A1145$ in the super-sequence,  we add another parameter to the construction of their ID: a "\textit{duplicate-id}", which works on the same principle as "\textit{twin-id}".
Thus in $SPL_{i+1}$, we find the artefact $A1145_{\_2}$, a mandatory duplicate of $A1145$. 

With each integration, the number of artefacts in the SPL's ART only increases, allowing us to guarantee the old products' integrity because nothing is deleted.

\subsection{Variability Computation}
\label{sec:variability}

\begin{figure}
    \centering
    \ifgpce
    \includegraphics[width=0.43\textwidth]{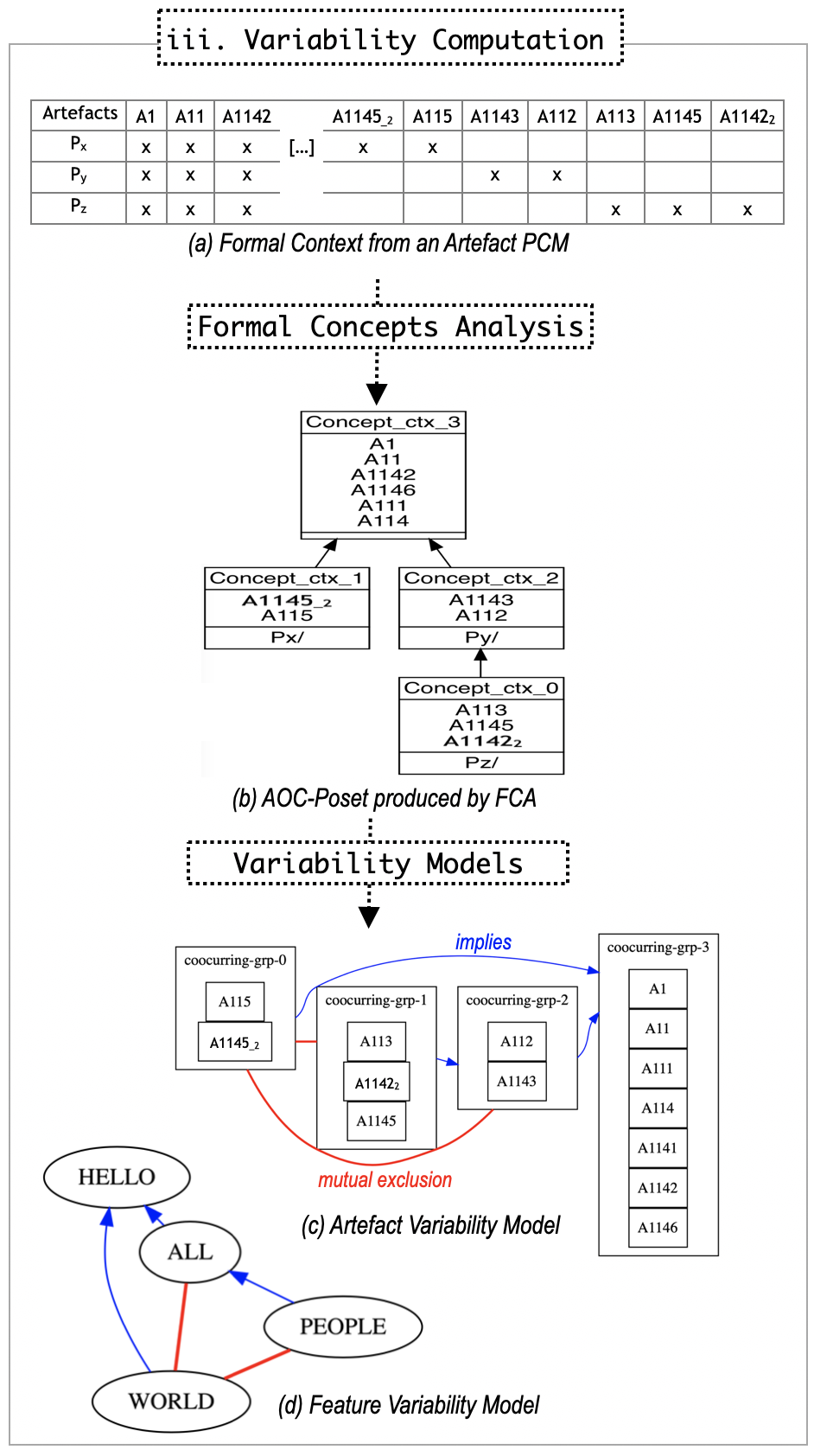}
    \else
    \fbox{\includegraphics[width=0.7\textwidth]{Integration/variability.png}}
    \fi
    \caption{The variability computation process run on $SPL_{i+1}$ implementation obtain in Fig. \ref{fig:identification}}
    \label{fig:variability}
\end{figure}

After the artefact acquisition, we now explain how the variability is managed, and thus our solution to problem 4. 
We apply Formal Concept Analysis (FCA) \cite{ganter_formal_1999} to determine the variability of the artefacts and the features and their constraints. 

\subsubsection{Formal Concept Analysis}
FCA works on a formal context, which is a set of associations between objects and attributes. 
Each object is described by a set of attributes that characterize it.
For instance, the software products can be the \textit{objects}, each of them having a set of artefacts as \textit{attributes}. 
FCA builds a set of concepts from a formal context. 
A concept is a maximum grouping of objects that share a minimum set of attributes.
A concept consists of an intention and an extension.
The concept's intention gives the set of its attributes, and its extension gives the set of objects that share these particular attributes.
Moreover, FCA establishes a partial order on these concepts, where a concept \(A\) is superior to a concept \(B\) if the attributes of the objects in \(B\) are in \(A\), or vice versa if the objects with the attributes in \(A\) are in \(B\). 
The partial order forms a hierarchy on these concepts, which allows building a lattice of concepts, also called an \textit{AOC-Poset}.

\subsubsection{Building a Formal Concept}
We start with the variability of an SPL's artefacts. 
Thanks to the unique IDs of the artefacts, we can create a formal context where we guarantee that each ID represents a specific artefact in the products.
To create this formal context, we assign each product to a list of the artefact ID, corresponding to the artefact identified in this product. 
We call this list an \textit{artefact configuration}.
A set of all artefact configuration forms a Product Configurations Matrix (PCM). 
A PCM has for columns the set of all the IDs in the SPL and for rows are the set of products, represented by their names.
This PCM is the formal context upon which we apply FCA, like shown in \textit{(a)} of Figure \ref{fig:variability}. 
FCA then builds an AOC-Poset from the PCM formal context, as displayed in part $(b)$ of the figure.

\subsubsection{Extracting the Variability}
The variability and the constraints are extracted from the lattice. 
This technique is adapted from the one presented by Carbonel et al. in their works \cite{carbonnel_clef_2019}, which they applied on a set of features.  
Here we apply it on both a set of artefacts and after a set of features.
The lattice first allows us to extract the binary variability (i.e. common or variable): the common artefacts are the ones in the top concept, here concept\_3. All the other artefacts are the variable ones.  
Next, we extract the following constraints among the artefacts : 
i) the implication, where selecting any artefact on concept\_0 will require selecting the artefact of concept\_2;
ii) the mutual exclusion, where the artefacts of concept\_1 and concept\_0 cannot be selected at the same time in the same product;
And iii) the co-occurrence, where the artefacts inside each concept must always be selected together. 
The algorithms used to extract these constraints are described in \cite{al-msiedeen_reverse_2014, carbonnel_clef_2019, carbonnel_analyse_2018}.

\subsubsection{Building the variability models}
These three constraints are used to create our Artefact Variability Model (AVM), see (c) in Figure \ref{fig:variability}, for a \(SPL_i\) at a given iteration \(i\) .
The AVM can assist developers during maintenance tasks, to give them more information about the variability and dependencies among their artefacts.
However, generating a new product requires a variability model of a more abstract level. 
Since the AVM is at an artefact level, it can be too complex to visualize as the number of artefacts grows.
Moreover, it is not practical to ask developers to select a list of IDs to be included in a new product. 
For these reasons, we also produce a Feature Variability Model (FVM), depicted in (d).
To build an FVM, we apply the same analysis with FCA as for the AVM. 
Except that this time, the formal context has for the attributes the feature names, and the objects remain the names of the products.
The FVM representation is analogous to the AVM, with the nodes being the feature instead of the artefacts. 

% variability model feature 
Equipped with an FVM, developers can visualize their features' variability and select them to generate a new product. 
However, this selection is only possible if we locate the actual feature implementation among the set of artefacts.

\subsubsection{Toward an automated Feature Location}
\label{sec:feature-location}

\begin{figure}
    \centering
    \ifgpce
    \includegraphics[width=0.463\textwidth]{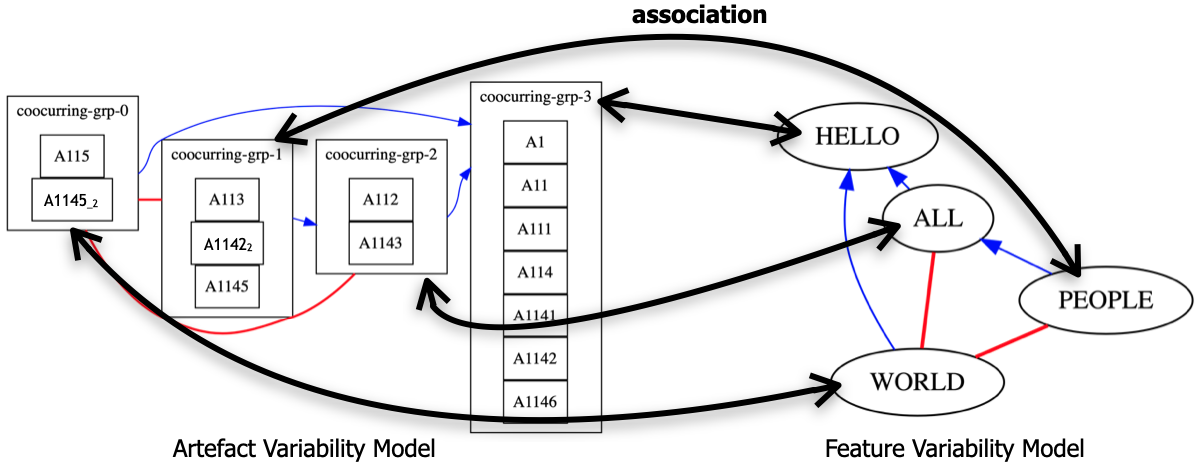}
    \else
    \fbox{\includegraphics[width=1\textwidth]{Integration/possible-FL.png}}
    \fi
    \caption{A Possible Feature Location based with isiSPL}
    \label{fig:possibleFL}
\end{figure}

\paragraph{A semi-automated Feature Location.}
For this first iteration of isiSPL, we have yet to define a fully automated process that embeds the Feature Location (FL). 
Thus, we consider that it is up to the developers to associate the set of artefacts with the set of features. 
However, we assist this task by proposing to go through the variability models. 
As presented in our last example, the AVM and the FVM present the variability of an SPL for the artefacts and features, which the developers can use to facilitate the association between a feature and its artefact group. 
Concretely, isiSPL allows the developers to make this association manually (see black arrows in Figure \ref{fig:possibleFL}) by letting the developers select which group of artefacts should be associated with features or feature interactions.

\paragraph{Perspectives to an automation.}
The FL must allow the automatic association between the elements of the AVM and the FVM. 
%A first idea would be to consider a correspondence between the nodes of two VMs.
As can be seen in Figure \ref{fig:possibleFL}, the two VMs are essentially the same graphs: their nodes are different, but they have the same number of nodes and the same edges between these nodes. 
The first idea would be to propose an ad hoc traversal of the two VMs by associating their corresponding nodes. 
When applied to our example, the cooccurring group $grp-3$ is mapped the $HELLO$ feature, as they are both the "common" element (implied by all others) in their respective graph. 
Following the implications, we can associate the rest of the groups: e.g. $grp-2$ with the feature $ALL$, because these nodes imply the common node, have a mutual exclusion, and are the conclusions of an implication. 
In the end, we obtain an association for each artefact group and each feature, as in Figure \ref{fig:possibleFL}

However, we have not implemented this technique because it does not work when there are more nodes in the AVM than in the FVM.
We noticed that there were always more nodes in the AVM when the products implemented a feature interaction between several features of the FVM. 
%This was the case in the majority of the SPLs we studied (ArgoUML , Sudoku, GameOfLife \cite{meinicke_mastering_2017}, etc.).
Moreover, since a feature interaction is not named in the product configuration, it does not give rise to a "new feature", and therefore does not appear as such in the FVM.

The second idea is to use FCA to retrieve feature-to-artefact associations. 
Using FCA would allow the same technology to be used for variability analysis and Feature Location.
Moreover, we already have the correct entries to apply FCA since we build the formal contexts of the features and the artefacts. 
Furthermore, previous work by Xue et al. \cite{Xue_Xing_Jarzabek_2012} and Salman et al. \cite{Al-Msiedeen_Seriai_Huchard_Urtado_Vauttier_Salman_2013, Salman_Seriai_Dony_Al-Msiedeen_2012} has shown promising results in applying FCA to FL. 
Nevertheless, they do not cover the association between artefacts and feature interactions. 
Therefore, we leave this issue open for future work.
In the rest of the paper, we will consider that the FL has been performed by the developers, as in Figure \ref{fig:possibleFL}.

%% file: Generation/generation-v3.tex
\section{The Generation Phase}
\label{sec:generation}

In this section, we present our automated Generation Phase that builds an annotative representation of an SPL Implementation, while minimizing the \textit{if/else hell} situation mentioned in \textit{Problem 5}. 
We also present how this representation can be pair to our variability model to generate new products. 

\subsection{SPL Generation}

\begin{figure}
    \centering
    \ifgpce
    \includegraphics[width=0.463\textwidth]{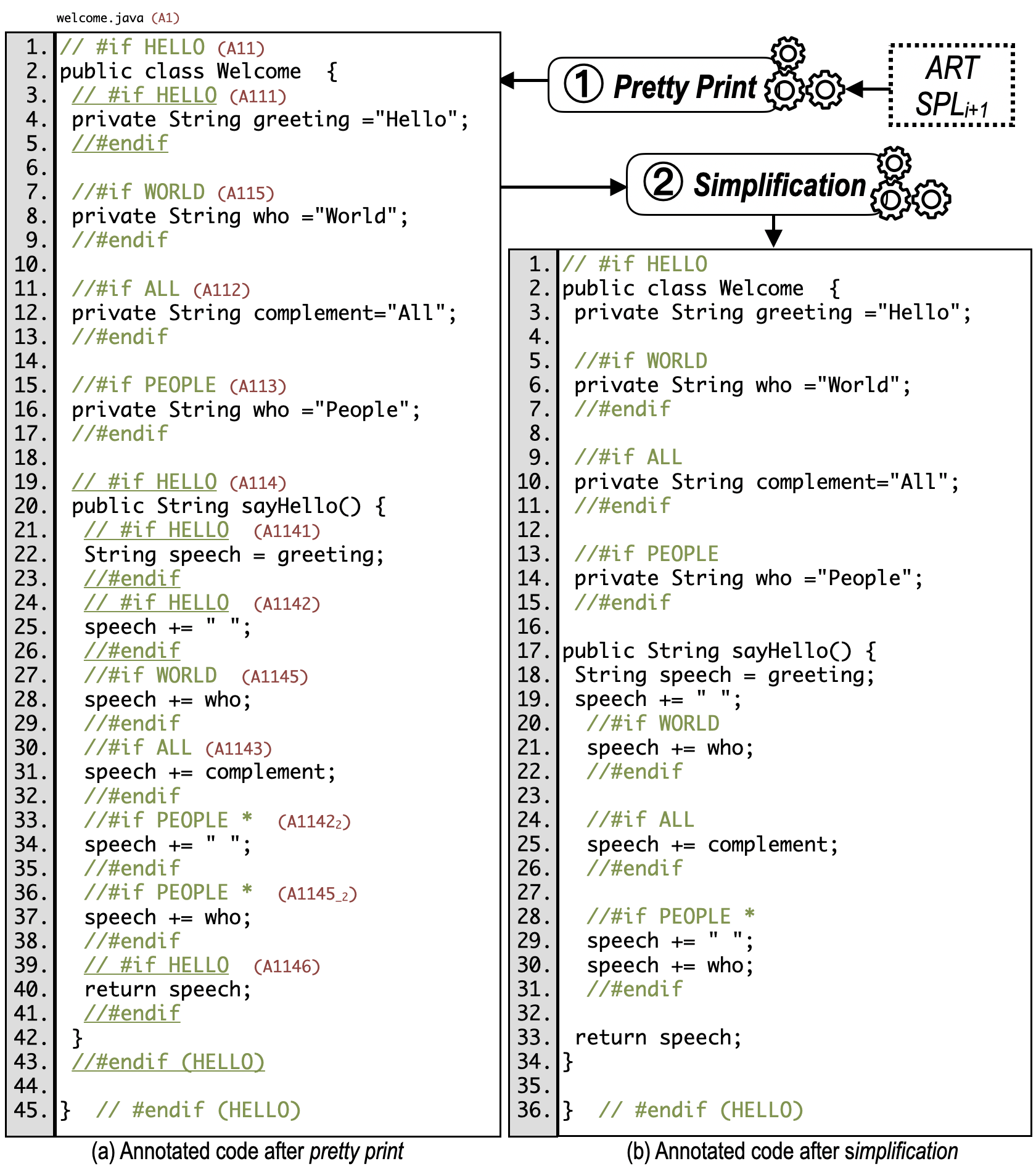}
    \else
    \fbox{\includegraphics[width=0.9\textwidth]{Generation/annotation-simplification}}
    \fi
    \caption{SPL Generation process, presented over the example of $SPL_{i+1}$ ART in Fig. \ref{fig:identification}.}
    \label{fig:prettyPrint}
\end{figure}

The C/C++ macros system inspires the Annotative representation of an SPL implementation. 
An annotation is declared over two comments, one for opening and one for closing it. We see an example of that the codes of figure \ref{fig:prettyPrint}. 
For instance in $(a)$, the annotation \textit{\#if HELLO} (line 3) declares the code in line 4 as part of the implementation of the feature \textit{Hello}.
As we see in this example, multiple annotations can be declared in the same file and sometimes nested.

To generate an annotated code for an SPL implementation, IsiSPL performs a \textit{pretty-print} operation on the ARTs. 
A \textit{pretty-print} transforms the content of an ART into a readable source code file while respecting the syntax and structure of the targeted language. 
For each ART, a code file is created at its corresponding path and an in-depth visit of this ART prints each artefact. 
Thus statements appear in the correct order since we print them in the order obtained with the super-sequence. 
The code of an optional artefact is surrounded by an annotation, which shows the names of the features traced to it during the Feature Location step. 
In Figure \ref{fig:prettyPrint} (a), we pretty-print the ART obtained at the end of figure \ref{fig:identification}, representing the file \textit{Welcome.java}. 
We add as a visual aid the corresponding artefact ID from the ART next to the annotation. However, we see that a naive pretty-print that includes all annotations makes the code rather difficult to read. 
It is the typical case of the intrinsic \textit{if/else hell} problem of the annotative approach, as explained in problem 5. 

In order to diminish the effect of the problem over the software comprehension, our solution is to perform a \textit{simplification} that removes redundant annotations from the printout:  

Simplification $S_1$ removes redundant nested annotations: for instance, all the annotations underlined \textit{\#if HELLO} inside the class \textit{Welcome} declaration are redundant, because the class itself is annotated by \textit{\#if HELLO}. 

Simplification $S_2$ merged multiple consecutive annotations that are the same : 
e.g. the two annotations line 33 and 36 in $(a)$ (marked with $(*)$), are redundant because there are consecutive and both annotated with \textit{PEOPLE}. Thus they are regroup into one annotation \textit{\#if PEOPLE} in $(b)$. 
After the simplification, we obtain a code that has signification fewer actual Lines of Code (LoC), and with fewer annotations (from 40 LoC in $(a)$ to 28 LoC in $(b)$).
As a result, the code is smaller and thus more comprehensible for the developers.

We should also mention that thanks to our fine artefact granularity (down to a statement-level), we can make annotations that contain few LoC.
This allows us to have small portions of annotated code and thus to factorize more codes. 

%Generation du produit
\subsection{Product Generation}
\label{sec:productgeneration}

Having generated the annotated code, the generation of a new product is a straightforward process. 
It essentially comes down to visit the annotated code and keep certain annotations, and remove others. 
These kept annotations are selected depending on the features selected by the developers. 
One consequence is that the quality of the generated code depends directly on the quality of the annotated code; if the latter is badly constructed (i.e. syntax error during generation, inversion of orders in the statements, forgotten artefacts), then the product code will be as well.
Moreover, if the product is correctly produced, it has the advantage of having its code being the same as the one developers see in the SPL implementation. 

Furthermore, because the generated code is from the SPL implementation code, it is also the same as the code of the integrated products. 
Thus, the developers who will manipulate this product may already be familiar with it, facilitating the development process.  
It should be mentioned that this code has no particularity other than being made from reusing code elements from other products.  
Therefore, developers can import the product into an IDE of their choice, and do not need specific SPLE training to develop it. 

Sometimes, through their selection of features using the FVM, developers may want to select features in mutual exclusion. 
For instance, the two features \textit{People} and \textit{All} from our first example never appear together inside one product. 
Thus the variability models will see them as mutually excluded. 
If the developers see the benefits of generating this configuration, isiSPL notify them by a warning message and generate the product.  
However, this code may contain some syntax errors and bugs, since the two features were never together in the same product.
We let the developers the responsibility to manage these bugs.

%% file: Validation/validation-v2.tex
\section{Validation}
\label{sec:validation}
\subsection{Research Question}
Our goal is to evaluate the automated aspects of isiSPL, independently from the Feature Location since it remains a semi-automated step in isiSPL. 
We want to validate that artefacts are correctly integrated into an SPL, and ensure that it does not impact the behavior of the previously integrated products.
To verify both aspects, we propose running the integration over a set of products and reproducing them through the generation phase.   
If the products can be restored without loss and without changing their behavior, then we can validate the integration and the generation.  
So to validate isiSPL, we seek to answer the following question:
\textbf{What is the success rate of the integration phase, when measuring it with the re-generation of integrated products?} 
Through this evaluation, we can get an indicator of the SPL Generation phase. 
Indeed, the annotated code generated by isiSPL is the basis on which the product generation is conducted (see \ref{sec:productgeneration}). 
Thus, evaluating products' reproduction also provides a good indicator of the correctness of an annotated code produced by isiSPL.

\subsection{Data Collection}

We have selected two datasets from the literature to perform our experiment.
The first one is \textit{ArgoUML-app} \footnote{\href{https://web.archive.org/web/20210623142746/https://github.com/marcusvnac/argouml-spl}{https://github.com/marcusvnac/argouml-spl}}, the main plugin of ArgoUML which is software for creating UML diagrams for computers. 
It is a large SPL of ten products that have between 1371 and 1502 files each. 
This SPL is used to test the scalability of isiSPL. 
The second is \textit{Sudoku} (available in the set of examples in FeatureIDE\cite{meinicke_mastering_2017} \footnote{\href{http://web.archive.org/web/20210624063757/https://featureide.github.io/}{https://featureide.github.io/}}), an SPL of 10 products all composed of 27 files. 
We include this one to see the performance over a smaller example. 

These datasets are often used as a benchmark, which will help to position our results. Moreover, they are all coded in Java. Finally, we have the original annotated code ArgoUML-app, which allows us to compare it manually to annotated code produced by isiSPL.

\subsection{Experimentation Protocol} 
\begin{figure}
    \centering
    \ifgpce
    \includegraphics[width=0.42\textwidth]{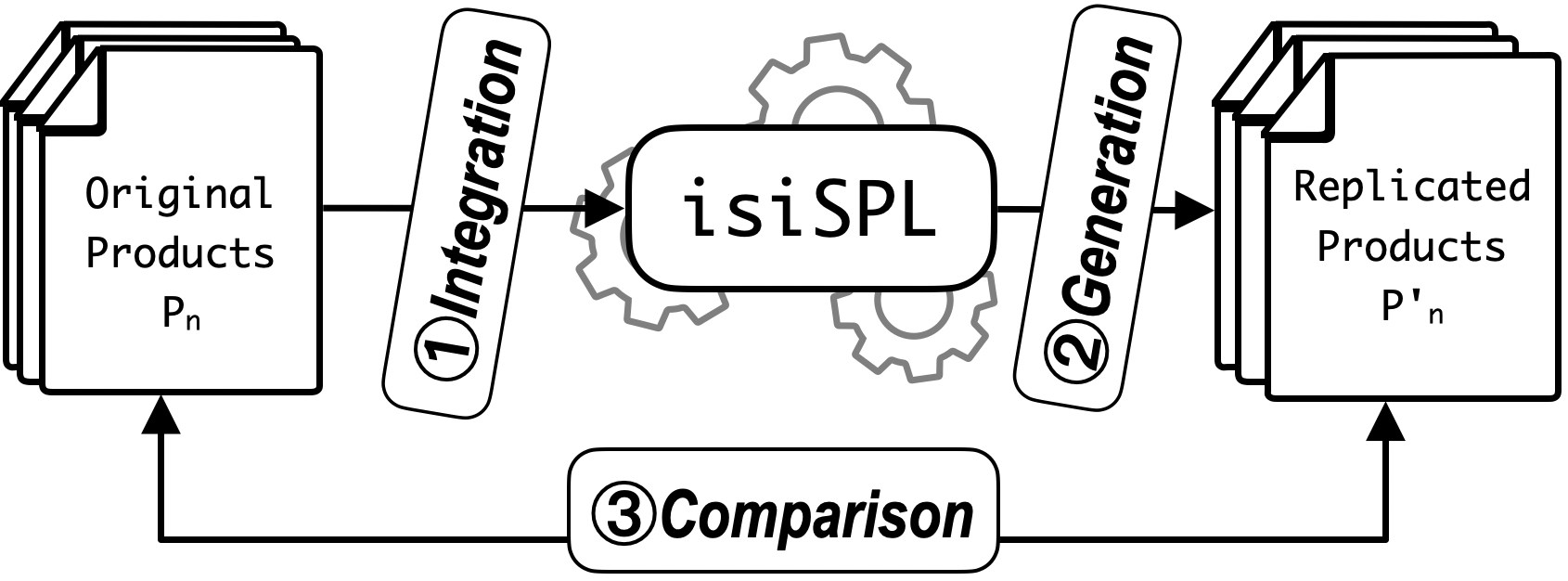}
    \else
    \fbox{\includegraphics[width=0.7\textwidth]{Validation/validation.png}}
    \fi
    \caption{Validation protocol}
    \label{fig:validation}
\end{figure}

% parler de l’outil en 1er. 
\subsubsection{isiSPL Prototype}
We made a Java implementation of isiSPL. 
This prototype uses different tools such as \textit{RCA-Explorer} to perform FCA \cite{dolques_rcaexplore_2019}, and \textit{CLEF} to extract the variability and constraints from the lattices into a variability models \cite{carbonnel_clef_2019}.
Our prototype includes the entire isiSPL's cycle, from the identification of the artefacts to the generation of the SPL implementation and its products. 
The annotation system of the SPL generation is based on the \textit{Antenna} annotation model \cite{Pleumann_Yadan_Wetterberg}, notably used in \textit{FeatureIDE} \cite{leich_tool_2005}. 
Our tool also generates variability models for the artefacts and features inside \textit{.dot} files. 
Our implementation runs on a 13 inch MacBook Pro, 2.3GHz Intel 4-Cores i5, 16Go RAM and 256 SSD. 

% traitement des produits 
\subsubsection{Protocol}
For each SPL, we execute the process described in Figure \ref{fig:validation}: 
(1) \textit{Integration}, the original products $P_o$ are integrated one by one by isiSPL, to simulate a iterative aspect. 
(2) \textit{Generation}, once all products are integrated, we reproduce the $P_g$ products from the new SPL built by isiSPL. 
(3) \textit{Comparison}, we compare the codes to count the errors introduced by isiSPL when it re-generates the products.  
% how to do the generation without including the localization
We voluntarily excluded the Feature Location, and so we designed a protocol for the Product Generation Phase to work without it. 
We modified the generation of the annotated code of an SPL, to replace the feature names in the annotations with artefacts ID. 
Thus, when generating a product, we present a list of artefact identifiers instead of a list of features. 
This Artefact Configuration is the same list obtained during the integration phase (see \ref{sec:integration}.  
This particular generation of products, independent of the feature location, is specific to this experiment and is not part of the normal isiSPL process.

To perform the comparison, we use \textit{Gumtree}, a \textit{diff} tool based on the comparison of the ASTs of two source code files \cite{falleri_fine-grained_2014}. 
Gumtree compares two AST over four criteria: insertions, deletions, updates and moves. 
We consider two products equal if Gumtree found no difference between them. 
However, we ignore some moves because, as the Product Generation pretty-prints from the ARTs, it is possible to see changes in the order of code elements (e.g. a change in the order of the declarations of the methods in a class). 
As we mentioned in the artefact identification (section \ref{sec:ident}), we only consider moves occurring at a statement level to have an impact upon a product behavior. 
Thus, we only count moves affecting the statements. 
To compare the \(P_g\) and \(P_o\), we look at the percentage of LoC affected by a difference. 
The differences are detected by Gumtree, which reports inserted, deleted, updated and moved elements in the code. 
We score these differences with $rep\_err$, in equation \ref{eq}.
The smaller the $rep\_err$, the closer the products are; and if it equals zero, then the products are identical.

\begin{equation}
rep\_err= \frac{total\; modified\; LoC\; in\; P_g}{total\; LoC\; in\; P_o} * 100 \label{eq}
\end{equation}

\subsection{Results}
For each tested dataset, no difference was detected by Gumtree, thus the $rep\_err$ was always equalled to zero. 
The result indicates that isiSPL integrates products automatically without damaging the previously integrated products. 
The total time to compute the integration and the generation was relatively low: Sudoku: 20 sec; 
%HealthWatcher: 5m46s;
and ArgoUML-app: 23min.
These performances are unmatched compared to what developers could do manually. 
We also run this experimentation protocol on other SPLs, and we have reported no differences between the originals products and the reproduced ones. 
This demonstrates that for these datasets, the integration and generation of isiSPL has a reproduction success rate of 100\%.
Thus no artefact was lost, and statements' sequences were preserved, keeping the product's behavior intact.   
We have included all our results and more in our Git repository \footnote{\url{https://github.com/isiSPL-results/SomeResultsRepo}}.

We also included the original annotated codes for as much SPL as we could to compare them with the one generated by isiSPL.
To produce these annotated codes with the feature names, we manually made the Feature Location using the semi-automated process mentioned in Section \ref{sec:feature-location}.
A manual comparison shows that the original and the generated SPL implementation tends to be similar, which indicates that isiSPL produces comparable SPL implementation of what developers can do.
However, this claim has yet to be validated with proper metrics to compare the two SPL implementations.

\subsection{Threats to validity} 
Our first threat concerned our current implementation of isiSPL in our prototype.
It runs on products coded in Java 1.7, which only validates our approach for this particular version of Java. 
Our prototype is still a work in progress; thus, the artefact identification has been simplified so that in-line or block comments were not taken into account.
We also manually reduced and modified the ArgoUML to ignore all the specificities of Java, in particular to ignore the case of single statement declarations without block (e.g. single-statement \textit{for}, \textit{if}, \textit{while}, etc.) 

Our entire dataset also represents a threat to validity. 
Since most products are taken from existing SPLs, they share a common architecture that facilitates their integration. 
To truly evaluate the relevance of isiSPL, we need to replicate development conditions closer to those faced by developers. 
However, we had difficulties finding industrial partners to carry on this experimentation, as it requires much time from the developers.

Furthermore, isiSPL assumes that products will share and maintain a common architecture throughout the life of an SPL.
However, the generation of new products from the SPL forces the developers to keep this common architecture present in their products.
If this is not the case, isiSPL can detect that a product is too different from the rest of the SPL and warn the developers. 

Finally, our evaluation does not yet examine the feature location step, as we only focus on the fully automated phase of our approach.

%% file: RelatedWork/relatedWork.tex
\section{Related Work}
\label{sec:relatedwork}

From an economic point of view, Buhrdorf et al. \cite{buhrdorf_salions_2003} study the adoption of a manual reactive approach in an industrial context. 
They observe a significant gain in resources and efforts when using the reactive approach compared to the proactive one.
Moreover, Schmid and Verlage \cite{schmid_economic_2002} propose an in-depth comparison of the reactive and the proactive approaches.
They also argue that the reactive approach can reduce costs and efforts, but only if the companies effectively manage it.
Indeed, they see that the reactive approach can be more expensive in some cases and that, in general, it is a riskier approach than a proactive one.
These studies are in line with our comments on the use of the reactive approach in the industry.  

In their systematic mapping \cite{assuncao_reengineering_2017}, the authors consider the re-engineering process of turning legacy products into an SPL in three generic phases: feature detection, variability analysis, and Transformation. The \textit{Transformation} phase is described as implementing variability in an SPL implementation using a simple mechanism such as the annotations. 
However, the authors mentioned a lack of study that focus on this phase. 
We believe that isiSPL can contribute to these researches, since its Generation phase corresponds to the "transformation" describe in \cite{assuncao_reengineering_2017}.

%Three studies share a common goal to ours. 
We also search in the literature for approaches that have comparable objectives to isiSPL.
%INCLINE
%\textbf{TK: dire qu'on propose une autre vue du problème}
Lillach et al. \cite{lillack_intention-based_2019} proposes an iterative approach to integrate a product variants inside a common \textit{configurable platform}, which is an open representation similar to an SPL implementation.
Their integration phase is semi-automatic and intention-based: developers must express their intents when introducing new artefacts or modifications to the configurable platform.
For example, a portion of code can be integrated with an \textit{keep\_as\_feature} intention, indicating that this code belongs to a specific feature.
Their tool, INCLINE, exposes the reusable artefacts with a binary variability (optional or common) and the feature-to-artefacts traces. 
However, unlike isiSPL, INCLINE does not compute the variability constraints of the artefacts automatically.
Moreover, the integration is semi-automatic, and intentions have to be expressed for each new artefacts integrated, while isiSPL requires no effort from the developers during the integration phase.

Fischer et al. \cite{fischer_ecco_2015, fischer_enhancing_2014, fischer_enhancing_2016} propose an extractive approach called \textit{ECCO}, to locate features inside product variants.
Like us, they propose to recover artefacts into an Artefact-Tree automatically and to trace them to features. 
However, our definitions of an ART differ, as ECCO builds several small artefact trees from the products (several per product) and only assemble them when it generates the code for a particular product. 
Furthermore, ECCO considers the code in the body of methods as \textit{raw lines}, each line of code being a new artefact. 
Whereas in isiSPL, the SPL stores only one Artefact-Tree per file, containing all the SPL's artefacts, and this ART is "pruned" to obtain the product's files. 
IsiSPL also regards each statement as individual artefacts inside the methods, not the line of code.
Moreover, ECCO has a black-box representation of an SPL implementation, with no annotated code or equivalent to allow its manipulation, which complicates the evolution process of an SPL. 
With isiSPL, we allow developers to directly manipulate an SPL implementation through its annotated code, which is intuitive and accessible for them.
ECCO computes a structural dependency graph from the source code analysis to extract the relationships between the different artefacts. 
Whereas isiSPL relies on FCA to recover the variability relationship, namely the implication, the mutual exclusion, the co-occurrence, and the common and optional artefacts (or feature). 
This means that isiSPL extracts these relationships independently of the source code. 
As our objectives are relatively similar, we hope to compare our approaches in the future.
%This comparison will focus on the feature location aspect, which is at the heart of ECCO's work and which we have not presented in this paper.

% Ziadi + But4Reuse
%Ziadi et al. 
Ziadi et al. \cite{ziadi_towards_2014} present their extractive tool, \textit{But4Reuse}.
This tool can extract a set of artefacts and features from a set of products, and associate them with traces with the feature location. 
But4Reuse has the advantage of working on various cases (different programming languages, images, musics...) thanks to its adapters system. 
However, their proposed representation of an SPL implementation somewhat naive because it only divides the code into different folders, each one being associated with an identified feature.
This makes maintainability efforts difficult since it hides the implementation of the variability.
Moreover, source code statements are not considered as artefacts, which causes significant duplicates in the different folders of their representation.
Finally, it seems to us that the maintenance and evolution can only be done by re-extracting the entire SPL after modifying the products.  
Thus, But4Reuse is more intended for product-managers than for developers.
In isiSPL, the variability is visible within the annotated code and its variability models,  allowing for the maintainability of an SPL implementation. 
%The maintainability is directly taken into account by modifying the SPL from the annotated code. 
Moreover, both product-managers and developers can take advantage of the feature variability model to generate new products and maintain their SPL.  

%similar technique
We also connect our work with the ones that share similar techniques. 
Al-Msiedeen et al. \cite{al-msiedeen_reverse_2014} and Carbonel et al \cite{carbonnel_clef_2019} also apply FCA to recover the constraints of mutual exclusion, implication and co-occurrence at the feature level.
We apply a similar technique at an artefact level in our work, considering the artefacts as our attributes instead. 
This allows us to build our variability model for the artefact, which is used to verify the product's configuration during the generation. 
Both \cite{al-msiedeen_reverse_2014, carbonnel_clef_2019} have proposed techniques to create a feature-model in a semi-automatic way. 
In future works, we want to focus on producing a feature-model for the developers and product managers.

%% file: Conclusion/conclusion-en.tex
\section{Conclusion and Future Works}
This paper proposes isiSPL, a reactive approach to facilitate SPL adoption in the software industry. 
IsiSPL automates the integration and generation phases in a non-invasive development cycle. 
It also proposes a concrete representation of an SPL implementation in the form of an annotated code, allowing the maintenance of an SPL by the developers.
In our validation, we show the efficiency of an isiSPL prototype to integrate and reproduce products.
The product reproduction is 100\% efficient, demonstrating that no loss nor error was introduced during the Integration phase. 

We want to focus on two main topics for our future works: the feature location and the industrial application of isiSPL. 
The Feature Location is crucial to automatizing the isiSPL entirely and transitioning from our proposed variability models toward the generation of a Feature-Model.
For the industrial part, we want to focus on evaluating isiSPL in an industrial environment, where we would evaluate its adoption and usefulness.
In particular, we want to compare the annotated codes and variability models to the ones built manually; and study if isiSPL is a viable approach in a real case scenario. 
Finally, we believe that transforming new product requirements in terms of features may be difficult for developers.
So in future work, we also want to explore ways to assist developers in this task and to include it inside isiSPL.